\documentclass[12pt]{article}
\usepackage[reqno]{amsmath}
\usepackage{bbm}
\usepackage{epsfig}
\usepackage{array}
\usepackage{float}
\usepackage{bm}

\usepackage{a4}

\usepackage{a4wide}
\usepackage{wasysym}

\def\ra{\rightarrow}
\def\be{\begin{equation}}
\def\ee{\end{equation}}
\def\gs{\mathrel{
   \rlap{\raise 0.511ex \hbox{$>$}}{\lower 0.511ex \hbox{$\sim$}}}}
\def\ls{\mathrel{
   \rlap{\raise 0.511ex \hbox{$<$}}{\lower 0.511ex \hbox{$\sim$}}}}

\newcommand{\ba}{\begin{array}{c}}
\newcommand{\baz}{\begin{array}{cc}}
\newcommand{\bad}{\begin{array}{ccc}}
\newcommand{\bea}{\begin{equation} \begin{array}{c}}
\newcommand{\eea}{ \end{array} \end{equation}}
\newcommand{\ea}{\end{array}}
\newcommand{\D}{\displaystyle}
\newcommand{\dms}{\mbox{$\Delta m^2_{\odot}$}}
\newcommand{\dma}{\mbox{$\Delta m^2_{\rm A}$}}
\newcommand{\meff}{\mbox{$\langle m \rangle$}}

\newcommand{\diag}{\ensuremath{\mathrm{diag}}}

\begin{document}

\title{\vspace{-2cm}
\hfill {\small TUM--HEP--631/06}\\
\vspace{-0.3cm}
\hfill {\small hep--ph/0605231} 
\vskip 0.6cm
\bf  Breaking of $L_\mu - L_\tau$ Flavor Symmetry, 
Lepton Flavor Violation and Leptogenesis 
}
\author{
Toshihiko Ota\thanks{email: \tt toshi@ph.tum.de}~~~~and~~
Werner Rodejohann\thanks{email: \tt werner.rodejohann@ph.tum.de}
\\\\
{\normalsize \it Physik--Department, Technische Universit\"at M\"unchen,}\\
{\normalsize \it  James--Franck--Strasse, D--85748 Garching, Germany}
}
\date{}
\maketitle
\thispagestyle{empty}
\vspace{-0.8cm}
\begin{abstract}
 
\noindent 
A supersymmetric see-saw model obeying the flavor 
symmetry $L_\mu - L_\tau$, 
which naturally predicts quasi-degenerate neutrinos,  
is investigated. Breaking of the symmetry is introduced in the 
Dirac mass matrix because it is the most economic choice in the 
sense that all interesting low and high energy phenomenology 
is made possible: 
we analyze the predictions for the low energy neutrino observables, 
for leptogenesis and for lepton flavor violating decays 
such as $\mu \to e \gamma$, 
where the SPS benchmark points for the SUSY parameters are 
used. 
It is outlined how these 
decays in connection with the requirement of successful leptogenesis and 
with correlations between the 
neutrino observables depend on the way the symmetry is broken. 

\end{abstract}

\newpage

\section{Introduction}
Explanation of the peculiar neutrino mass and mixing schemes is 
one of the most interesting tasks of particle physics. 
Motivated by spectacular experimental results, 
a very large number of models has been proposed in recent years \cite{revs}. 
Typically, the see-saw mechanism \cite{seesaw} is the starting 
point of most analyzes:
\be
m_\nu = - m_D^T \, M_R^{-1} \, m_D~,
\ee
where $M_R$ is the mass matrix of three heavy Majorana neutrinos 
$N_{1,2,3}$ and $m_D$ is a Dirac mass matrix resulting from 
the coupling of the Higgs doublet to the lepton doublets and 
the $N_i$. 
The light neutrino mass matrix  
$m_\nu$ is diagonalized by the PMNS matrix $U$ defined via 
\be \label{eq:def}
U^T \, m_\nu \, U = \diag(m_1,m_2,m_3)~. 
\ee
It can be parametrized as  
\begin{equation}
 U=\left(
 \begin{array}{ccc}
 c_{12} \, c_{13} & s_{12}\, c_{13} & s_{13}\, e^{-i \delta}\\
 -c_{23}\, s_{12}-s_{23}\, s_{13}\, c_{12}\, e^{i \delta} &
 c_{23}\, c_{12}-s_{23}\, s_{13}\, s_{12}\, e^{i \delta} & s_{23}\, c_{13}\\
 s_{23}\, s_{12}-\, c_{23}\, s_{13}\, c_{12}\, e^{i \delta} &
 -s_{23}\, c_{12}-c_{23}\, s_{13}\, s_{12}\, e^{i \delta} & c_{23}\, c_{13}
 \end{array}
 \right) \, P ~,
\end{equation}
where $P = \diag(1,e^{i \alpha},e^{i (\beta + \delta)})$ 
and $c_{ij}$, $s_{ij}$ are 
defined as $\cos\theta_{ij}$ and $\sin\theta_{ij}$, respectively. 
By making assumptions for the unknown neutrino parameters 
(in particular the mass scale, ordering and phases), 
one can reconstruct $m_\nu$ with the help of our current knowledge of $U$ and 
the mass differences \cite{thomas}. 
Atmospheric neutrino mixing is close to maximal, $\theta_{23} \simeq \pi/4$, 
and corresponds to a large $\dma = |m_3^2 - m_1^2| 
\simeq 2.5 \cdot 10^{-3}$ eV$^2$, whereas solar neutrino mixing is 
large but non-maximal, $\theta_{12} \simeq \pi/5$,  
and corresponds to a small $\dms = m_2^2 - m_1^2 \simeq 8 \cdot 10^{-5}$ 
eV$^2$. The third mixing angle is known to be smaller than 
roughly $\pi/15$. Nothing is known about the mass scale, the mass 
ordering (sign of \dma) and the phases. 

Several interesting hints towards the structure of $m_\nu$ can thereby be 
obtained, for instance the possibility of a $\mu$--$\tau$ exchange symmetry 
\cite{mutau,CR}. 
One other possible point of view is that a simple Abelian $U(1)$ 
symmetry is directly or effectively working on $m_\nu$. 
Conservation of a flavor charge 
is implied by the conservation of this $U(1)$ and well-known cases 
are $L_e$ \cite{le} and $L_e - L_\mu - L_\tau$ \cite{lelmlt}, 
which lead to a normal ($m_3^2 \gg m_{1,2}^2$) and 
inverted ($m_2^2 \simeq m_1^2 \gg m_3^2$) mass hierarchy, 
respectively. Recently the case 
$L_\mu - L_\tau$ has been found to be also possible \cite{CR}. 
A low energy mass matrix conserving $L_\mu - L_\tau$ has the form 
\be \label{eq:lmlt}
m_\nu =  m_0 \, 
\left( 
\bad 
a & 0 & 0 \\[0.3cm]
\cdot & 0 & b \\[0.3cm]
\cdot & \cdot & 0 
\ea 
\right) 
\ee
and for $a \simeq b$ one is lead to quasi-degenerate light 
neutrinos, i.e., masses $m_3 \simeq m_2 \simeq m_1 \equiv m_0 \ls $ eV 
much larger than the mass splittings. 
The neutrino mixing as predicted by the above matrix 
corresponds to $\theta_{13}=\theta_{12}=0$ and $\theta_{23}=\pi/4$, 
which reflects the $\mu$--$\tau$ symmetry inherent in a matrix conserving 
$L_\mu - L_\tau$.  We remark here that $L_e$ and $L_e - L_\mu - L_\tau$ 
do not possess $\mu$--$\tau$ symmetry. 
Note further that besides $\theta_{12}=0$ also $\dma = 0$ holds.
However, due to the quasi-degeneracy of the neutrinos, 
breaking of the symmetry with small parameters allows to easily 
overcome these shortcomings \cite{CR,RS,indian}.\\ 

The flavor symmetry $L_\mu - L_\tau$ 
can be incorporated in a see-saw model \cite{CR}. 
The relevant Lagrangian reads 
\be \label{eq:L} 
-{\cal{L}} = 
 \overline{N}_i \, (m_D)_{i \alpha} \, (\nu_\alpha)_L + 
\frac{1}{2}  \, \overline{N}_i \, (M_R)_{ij} \, N_j^c + h.c. 
\ee
Here the superscript $^c$ denotes charge conjugation. 
The charge assignment of the particles under $L_\mu - L_\tau$ is 
given in Table \ref{tab:charges}.
\begin{table}[t]
\begin{center}
\begin{tabular}{|c|c|c|c|c|c|c|c|} 
\hline
& $(\nu_e, e)_L$ &  $(\nu_\mu, \mu)_L$ &  
$(\nu_\tau, \tau)_L$ & $N_1, e_R$ & $N_2, \mu_R$ & $N_3, \tau_R$ & $\Phi$ 
\\ \hline \hline 
$L_\mu - L_\tau$ & 0 & 1 & $-1$ & 0 & 1 & $-1$ & 0 \\ \hline
\end{tabular}
\caption{\label{tab:charges}Particle content and charge under the $U(1)$ 
symmetry corresponding to $L_\mu - L_\tau$. Here 
$\Phi$ denotes the Higgs-doublet, which is responsible 
for the Dirac mass term.}
\end{center}
\end{table}
As a consequence, the charged lepton mass matrix is diagonal and 
in terms of mass matrices, we have 
(with $v_u = v \, \sin \beta$, $v = 174$ GeV, 
$\tan \beta$ the ratio of the up- and down-type Higgs doublets and $M$ 
the high mass scale of the heavy singlets)   
\be \label{eq:mdMR}
m_D = v_u \, 
\left( 
\bad 
a & 0 & 0 \\[0.2cm]
0 & b & 0 \\[0.2cm]
0 & 0 & d 
\ea 
\right)
\mbox{ and } 
M_R = M \, 
\left( 
\bad 
X 
& 0 & 0 \\[0.2cm]
\cdot & 0 & Y 
\\[0.2cm]
\cdot & \cdot & 0 
\ea 
\right)~.
\ee
One eigenvalue of $M_R$ has a mass $M \, X$ and there is a Pseudo-Dirac pair 
with masses $\pm M\, Y$. The low energy neutrino mass matrix is given by 
\be \label{eq:see--saw} 
m_\nu = - m_D^T \, M_R^{-1} \, m_D
= 
- \frac{v_u^2}{M} 
\left(
\bad 
\D \frac{a^2
}{X} & \D 0 & \D 0 \\[0.3cm]\D
\D \cdot & \D 0 & \D \frac{b \, d
}{Y}  \\[0.3cm]
\D \cdot & \D \cdot & \D 0 
\ea 
\right)~.
\ee
Note that the form of $M_R$ corresponds to the form of $m_\nu$ from 
Eq.~(\ref{eq:lmlt}). The parameters $a,b,d,X,Y$ are allowed by the 
symmetry and are therefore naturally of order one. 
As mentioned above, we need to break the symmetry in order to reproduce 
a non-zero atmospheric mass squared difference and a non-zero solar 
neutrino mixing angle. 
In addition, as we will see, 
successful leptogenesis and the existence of Lepton Flavor Violating (LFV) 
charged lepton decays such as $\mu \ra e \gamma$  
also require breaking terms. 
The possibilities to break the symmetry are numerous: we can 
\begin{itemize}
\item break $L_\mu - L_\tau$ in the 
charged lepton sector. This will allow only the generation of 
$\theta_{12} \neq 0$ (note that large mixing has to be generated) 
and for LFV decays; 
\item break $L_\mu - L_\tau$ in $M_R$. This will allow only for 
$\theta_{12} \neq 0$, $\dma \neq 0$ and for leptogenesis. 
Breaking in $M_R$ has previously been analyzed in \cite{CR,RS,indian}; 
\item  break $L_\mu - L_\tau$ in $m_D$. This will allow for
$\theta_{12} \neq 0$, $\dma \neq 0$, for leptogenesis and for LFV decays.
\end{itemize}
We conclude that breaking $L_\mu - L_\tau$ in $m_D$ is the most 
economic choice when one wants to generate all interesting observables. 
Of course, one would expect breaking in all possible sectors, but this 
will lead to little predictivity. For the sake of definiteness, we 
therefore consider only breaking in $m_D$.\\

In the following we will consider the case that the entries in $M_R$ 
are complex and in $m_D$ are real. The heavy neutrino mass matrix is 
\be \label{eq:defVR}
\frac{M_R}{M} =  
\left( 
\bad 
X \, e^{i \omega}
& 0 & 0 \\[0.2cm]
\cdot & 0 & Y \, e^{i \phi}
\\[0.2cm]
\cdot & \cdot & 0 
\ea 
\right) = 
V_R^\ast \, M_R^{\rm diag} \, V_R^\dagger \equiv 
P_R \, \tilde{V}_R \, Q_R \, 
\left( 
\bad 
X & 0 & 0 \\[0.2cm]
\cdot & Y & 0 \\[0.2cm]
\cdot & \cdot & Y 
\ea 
\right)\, Q_R  \, \tilde{V}_R^T \, P_R~,
\ee
where we have defined 
\be
\tilde{V}_R = 
\left( 
\bad 
1 & 0 & 0 \\[0.2cm]
0 & \sqrt{\frac 12 } & \sqrt{\frac 12 }  \\[0.2cm]
0 & -\sqrt{\frac 12 } & \sqrt{\frac 12 }
\ea 
\right)~~,~
P_R = 
\left( 
\bad 
 e^{i \omega/2} & 0 & 0 \\[0.2cm]
0 & e^{i \phi} & 0 \\[0.2cm]
0 & 0 & 1
\ea 
\right) 
\mbox{ and } 
Q_R = 
\left( 
\bad 
 1 & 0 & 0 \\[0.2cm]
0 & i  & 0 \\[0.2cm]
0 & 0 & 1
\ea 
\right) ~.
\ee
For real entries in $M_R$ the matrix $P_R$ is the unit matrix. 
When the breaking of $L_\mu - L_\tau$ takes place 
only in $m_D$ we can quantify this as 
\be \label{eq:defmd}
m_D = v_u \, 
\left(
\bad
a & \epsilon_1 & \epsilon_2 \\[0.2cm]
\eta_1 & b & \epsilon_3 \\[0.2cm]
\eta_2 & \eta_3 & d
\ea
\right)~,
\ee
with $\epsilon_i, \eta_i \ll 1$. For symmetric $m_D$ it would hold that 
$\epsilon_i = \eta_i$. With real entries in the Dirac mass matrix, there 
is only one physical phase, namely $\omega - \phi$. Consequently, 
low and high energy $CP$ violation will be intimately related. 
We are therefore allowed to set $\phi$ to zero and keep only the 
phase $\omega$. 
In the remaining part of this Section we will give the relevant expressions 
for the general form of $m_D$ from Eq.~(\ref{eq:defmd}), before 
considering more minimal braking scenarios in the next Section. 

The parameters and the breaking are introduced at high scale. 
Consequently, radiative corrections, both below and in between the 
see-saw scales, can have impact on the results. It has been 
shown in Ref.~\cite{RS}, however, that in the see-saw model 
basing on $L_\mu - L_\tau$ typically only 
$\theta_{12}$ gets corrected and that $\theta_{23}$ and 
$|U_{e3}|$, on which we later focus on, witness only little effects. 
Moreover, the textures of the mass matrices do 
only slightly change, i.e., small perturbations (over which we will 
scan numerically) remain small. We therefore 
neglect radiative effects, which should 
be a good approximation for our purposes.\\

In supersymmetric frameworks with universal boundary conditions 
there is an important possibility to 
probe the see-saw parameters, namely lepton flavor violating decays 
of charged leptons \cite{LFV}. 
In the leading-log approximation one can obtain for the branching ratios 
of the decays $\mu \ra e \gamma$, $\tau \ra e \gamma$ and 
$\tau \ra \mu  \gamma$ the following formula \cite{LFV}: 
\begin{equation}
\text{B}(\ell_i \to \ell_j   \gamma) \simeq 
\frac{\alpha_{\text{em}}^3}{G_F^2 \, m_S^8 \, v_u^4}
\left|\frac{(3 m_0^2 + A_0^2) }{8\pi^2}\right|^2
\left| \left(\tilde{m}_D^\dagger \, L \, \tilde{m}_D \right)_{ij}
\right|^2 \, \tan^2\beta\;, 
\label{eq_ijg}
\end{equation}
where $\ell_i = e, \mu, \tau$ for $i = 1,2,3$. 
Here $m_0$ is the universal scalar mass, $A_0$ the universal trilinear 
coupling parameter, $m_S$ represents a SUSY particle mass and 
$L = \ln \delta_{ij} \, M_i/M_X $, with $M_i$ the heavy Majorana masses and 
$M_X = 2 \cdot 10^{16}$ GeV. 
The branching ratios have to be evaluated in the basis in which the 
heavy Majorana neutrinos are real and diagonal. To get into this basis 
we have to rotate $m_D$ to obtain $\tilde{m}_D$. Having defined the 
diagonalization of $M_R$ in Eq.~(\ref{eq:defVR}) as 
$M_R = V_R^\ast \, M_R^{\rm diag} \, V_R^\dagger$, then 
\be
m_D \ra \tilde{m}_D = V_R^T \, m_D~.
\ee
At 90\% C.L., the current limit on the 
branching ratio of $B(\mu \to e \gamma)$ is 
$1.2 \cdot 10^{-11}$ \cite{mega} and future improvement by two 
orders of magnitude is expected \cite{psi}.
In most of the relevant soft 
SUSY breaking parameter space, the expression 
$m_S^8\simeq 0.5~m_0^2~m_{1/2}^2~(m_0^2 + 0.6 ~m_{1/2}^2)^2$, 
with $m_{1/2}$ being the universal gaugino mass, 
is an excellent approximation to the results obtained in a full
renormalization group analysis \cite{PPTY03}. Apparently, the 
branching ratios depend crucially on the SUSY masses. 
We choose here to use as examples the 
SPS benchmark points from Ref.~\cite{SPS} as given in Table \ref{tab:SPS}.

Denoting 
$(3 m_0^2 + A_0^2)^2/m_S^8$ with $1/\tilde m_S^4$, we can write 
\be \label{eq:estimate}
\text{B}(\mu \to e  \gamma) \simeq 1.2 \cdot 10^{-9} \, 
\left(\frac{200 \, \rm GeV}{\tilde m_S} \right)^4 \, 
\left| \left(\tilde{m}_D^\dagger \, L \, \tilde{m}_D \right)_{21}
\right|^2 \frac{1}{v_u^4} \, 
\, \tan^2\beta~,
\ee 
which has to be smaller than $10^{-11}$. As we will see below, this 
can constrain the way $L_\mu - L_\tau$ should be broken. 

\begin{table}
\begin{center}
\begin{tabular}{|c|c|c|c|c|} \hline 
Point & $m_0$ & $m_{1/2}$ & $A_0$ & $\tan \beta$ \\ \hline \hline 
1a    & 100   & 250       & $-100$  & 10 \\ \hline 
1b    & 200   & 400       & 0       & 30 \\ \hline 
2     & 1450  & 300       & 0       & 10 \\ \hline 
3     & 90    & 400       & 0       & 10 \\ \hline 
4     & 400   & 300       & 0       & 50 \\ \hline 
5     & 150   & 300       & $-1000$ & 5 \\ \hline 
\end{tabular}
\end{center}
\caption{\label{tab:SPS}SPS benchmark values for the mSUGRA parameters 
according to Ref.~\cite{SPS}. The values of $m_0$, 
$m_{1/2}$ and $A_0$ are in GeV. }
\end{table}

It proves useful to consider
also the double-ratios 
\be
 \D
\frac{\text{B}(\mu \to e + \gamma)}
  {\text{B}(\tau \to e + \gamma)}
\simeq 
\frac{\left| (\tilde{m}_D^\dagger \, L \, \tilde{m}_D )_{21}
\right|^2}{\left| (\tilde{m}_D^\dagger \, L \, \tilde{m}_D )_{31}
\right|^2}~~\mbox{ and } ~~
 \D
  \frac{\text{B}(\mu \to e + \gamma)} {\text{B}(\tau \to \mu +
    \gamma)}
\simeq 
\frac{\left| (\tilde{m}_D^\dagger \, L \, \tilde{m}_D )_{21}
\right|^2}{\left| (\tilde{m}_D^\dagger \, L \, \tilde{m}_D )_{32}
\right|^2}\;,
\label{DoubleR}
\ee
which are essentially independent on the SUSY parameters.  

With the most general breaking structure in $m_D$ given in 
Eq.~(\ref{eq:defmd}) and with using $L_3 = L_2$, 
the off-diagonal entries of $\tilde{m}_D^\dagger \, L \, \tilde{m}_D$ read 
\bea \label{eq:mdtmd}
(\tilde{m}_D^\dagger \, L \, \tilde{m}_D)_{12} = 
a \, \epsilon_1  \, L_1 + (b \, \eta_1 + \eta_2 \, \eta_3) \, L_2~~,~\\[0.3cm]
(\tilde{m}_D^\dagger \, L \, \tilde{m}_D)_{13} = 
a \, \epsilon_2  \, L_1 
+ (d \, \eta_2 + \epsilon_3 \, \eta_1) \, L_2 ~~,~\\[0.3cm]
(\tilde{m}_D^\dagger \, L \, \tilde{m}_D)_{23} = 
\epsilon_1 \, \epsilon_2 \, L_1 + (b \, \epsilon_3 + d \, \eta_3) \, L_2
~.
\eea
If $L_\mu - L_\tau$ would be broken only 
in the heavy neutrino sector (as in Refs.~\cite{CR,RS,indian}), then 
$\tilde{m}_D^\dagger \, L \, \tilde{m}_D $ would be diagonal and the 
decays would be extremely suppressed. If we break $L_\mu - L_\tau$ 
only in the charged lepton sector, then $\tilde m_D = m_D \, U_\ell$, 
where $U_\ell$ diagonalizes the (now non-diagonal) charged lepton 
mass matrix. In this case  
$\tilde{m}_D^\dagger \, L \, \tilde{m}_D$ will have off-diagonal entries, 
but leptogenesis, to be discussed in the next paragraph, 
will not be possible.\\ 

Another very helpful and interesting aspect of see-saw models is 
the possibility to generate the baryon asymmetry of the Universe  
with the help of the leptogenesis mechanism \cite{lepto}. 
In the case of thermal leptogenesis 
the baryon asymmetry is given by (for a review see, e.g., \cite{bumu})
\begin{equation}
\eta_B  = \frac{n_B}{n_\gamma} 
\simeq - 1.04 \cdot 10^{-2}\, \kappa\, \varepsilon_1\;,
\label{YBobsth}
\end{equation}
where $\varepsilon_1$ is the $CP$-violating asymmetry in the decay of the
lightest right-handed Majorana neutrino $N_1$ having the mass $M_1$, and
$\kappa$ is an efficiency factor calculated by solving the Boltzmann
equations. A simple approximate 
expression for the efficiency factor $\kappa$ in the case of thermal
leptogenesis was given in~\cite{CERN04}:
\begin{align} \label{eq:kappa}
\frac{1}{\kappa} \simeq \frac{3.3\cdot 10^{-3}~\text{eV}}{\widetilde{m}_1}
+\left(\frac{\widetilde{m}_1}{0.55\cdot 10^{-3}~\text{eV}}\right)^{1.16}\;,
\end{align}
where the important parameter $\widetilde{m}_1$ is given by 
\begin{align}
\widetilde{m}_1 \equiv 
\frac{ (\tilde{m}_D \, \tilde{m}_D^\dagger )_{11} }{M_1}\,.
\end{align}
The $CP$-violating decay asymmetry $\varepsilon_i$ has the form
(with $x_j = M_j^2 /M_i^2$): 
\be 
\varepsilon_i \D 
= \frac{1}{8 \pi \, v_u^2} \, 
\frac{1}{(\tilde{m}_D \, \tilde{m}_D^\dagger)_{ii}}  
\, \sum\limits_{j \neq i} 
{\rm Im} \left\{ \left(\tilde{m}_D \, \tilde{m}_D^\dagger 
\right)_{ji}^2 \right\} \, 
\sqrt{x_j} \, \left(
\frac{2}{1 - x_j} - \ln \left( \frac{1+x_j}{x_j} \right) 
 \right) 
~. \label{e1H}
\ee
For neutrinos close in mass the (self-energy) term proportional 
to $(1-x_j)^{-1}$ dominates \cite{YBreso}. 
It is important to note here that 
the Pseudo-Dirac pair of mass $Y \, M$ generates no 
decay asymmetry. The decay asymmetry is therefore generated by the decay 
of the neutrino with mass $M \, X$. 
It holds that 
$\tilde{m}_D \, \tilde{m}_D^\dagger 
= V_R^T \, m_D \, m_D^\dagger \, V_R^\ast$. In case of $X \sim Y$ we 
have\footnote{For extremely degenerate and therefore fine-tuned 
heavy neutrinos one should use a different 
formula \cite{YBreso}.} 
\be \label{eq:epsi}
\varepsilon_X \simeq - \frac{1}{4 \pi} 
\, \frac{1}{a^2 + \epsilon_1^2 + \epsilon_2^2} \, \frac{1}{Y/X - X/Y} \, 
\, 
2 \, (b \, \epsilon_1 + a \, \eta_1 + \epsilon_2 \, \epsilon_3 ) 
(d \, \epsilon_2 + a \, \eta_2 + \epsilon_1 \, \eta_3) 
\, \sin \omega 
\ee
and 
\be
\widetilde{m}_1 = \frac{v_u^2}{M}\,
\frac{ a^2 + \epsilon_1^2 + \epsilon_2^2 }{X}\,.
\label{tilm1}
\ee
Note that for no breaking of $L_\mu - L_\tau$ (i.e., $\eta_i = \epsilon_i=0$) 
the decay asymmetry vanishes. In addition, if we break the symmetry only 
in the charged lepton sector we would have no decay asymmetry either, 
because $\tilde{m}_D \, \tilde{m}_D^\dagger$ would remain diagonal. 

Numerically, $\eta_B$ should be given by $6 \cdot 10^{-10}$ \cite{YB}, 
where the small error is on the $5\%$ level. 
The formalism described above has however several sources of uncertainty. 
First, recall that expression (\ref{eq:kappa}) holds only for hierarchical 
heavy neutrinos. The wash-out effect of the neutrinos with mass 
$\pm Y \, M$ is therefore not properly taken into account. 
Second, it has recently been realized \cite{flavor} that 
flavor effects in leptogenesis can significantly affect the results. 
Taking these issues into account would require a thorough study 
and solution of the Boltzmann equations, which is surely beyond 
the scope of this letter. Instead, when we in the next Section 
calculate the baryon asymmetry for a specific breaking scenario, 
we consider the calculation as successful, when 
the result is $4 \cdot 10^{-10} \le \eta_B \le 8 \cdot 10^{-10}$, 
which is presumably still a very conservative range.

\section{Breaking of $L_\mu - L_\tau$: Concrete Examples}

Up to now we gave the relevant expressions for the Dirac mass matrix 
from Eq.~(\ref{eq:defmd}), i.e., we used the most general breaking 
scenario. 
With six arbitrary breaking parameters in $m_D$, however, 
there is little predictive power in 
what regards the observables and in order 
to make interesting statements more simplification is needed. 
We therefore turn to minimal breaking scenarios in the sense of having 
as few parameters as possible. 
To constrain the possibilities even more, we require the 
presence of both low and high energy $CP$ violation. 
If there is low energy $CP$ violation in oscillation experiments 
can be checked most easily by calculating the 
following invariant \cite{branco}, to which any $CP$ violation in 
neutrino oscillations has to be proportional: 
\bea \label{eq:jcp}
{\rm Im} \left\{ h_{12} \, h_{23} \, h_{31} \right\} = 
\Delta m^2_{21} \, \Delta m^2_{31} \, \Delta m^2_{32}~J_{CP}~, \\[0.2cm]
\mbox{where } 
h = m_\nu^\dagger\, m_\nu~\mbox{ and } 
J_{CP} = \frac 18 \, \sin 2 \theta_{12} \, \sin 2\theta_{23} \, 
\sin 2 \theta_{13} \, \cos \theta_{13} \, \sin \delta ~. 
\eea
With only one perturbative parameter in $m_D$, this expression always 
vanishes. Hence, we should analyze scenarios with two non-zero 
perturbations in $m_D$, for which there are 15 possibilities. 
Except for one case, low energy mass matrices with one 
or two zeros are generated. 
Only one of them is ruled out by neutrino data, 
namely when the 23 and 32 elements of $m_D$ are filled with non-zero 
entries. A low energy mass matrix with zeros in the 12 and 13 element 
would result, which can not reproduce the data \cite{2zeros}. 
More cases can be 
ruled out when we require the presence high energy $CP$ violation, i.e., 
leptogenesis. 
Recall that the decay asymmetry is proportional to 
$(b \, \epsilon_1 + a \, \eta_1 + \epsilon_2 \, \epsilon_3 ) 
(d \, \epsilon_2 + a \, \eta_2 + \epsilon_1 \, \eta_3) $. Asking this 
expression to be non-zero rules out 8 more cases, leaving us 
with 6 remaining ones. 
In what regards the results of the breaking scenarios, we are interested 
in particular in the branching ratios of 
the LFV decays, some of which will be forbidden by certain scenarios. 
We have summarized in Table \ref{tab:break} all 15 possibilities 
together with their predictions for the 
branching ratios of the LFV processes, for low energy $CP$ violation, 
for $\eta_B$, and with a correlation for 
the oscillation parameters as obtained in \cite{2zeros,MR}. 
All obtained cases with two zeros are 
only possible for quasi-degenerate neutrinos \cite{2zeros}. 
Cases with one zero entry 
are in general possible also for other allowed mass 
hierarchies \cite{MR}, but here we 
focus only on quasi-degenerate neutrinos. The one-zero 
matrices always come together with zero $\eta_B$ and are disregarded anyway. 
From the six cases allowing for a non-zero baryon asymmetry one case 
has no correlation for the low energy observables 
and all branching ratios are non-zero. 
Four cases generating two zeros in the low energy mass matrix have 
indistinguishable neutrino phenomenology, but differ in the 
predictions for the branching ratios, except for 2 cases which predict 
identical results.  
From the six matrices allowing for leptogenesis, five 
also predict the decay $\mu \to e \gamma$. 
We would like to remark here that some of the 15 possibilities 
have the amusing feature that  
{\it there is low energy $CP$ violation but no leptogenesis.} There are 
no cases in which it is the other way around. 
Note finally that the rate for neutrinoless double beta decay 
(which is proportional to the $ee$ element of $m_\nu$) 
is always non-zero.

Let us discuss one example in detail, namely the following form of 
$m_D$ and the resulting low energy mass matrix $m_\nu$: 
\be \label{eq:exam1}
m_D = v_u \, 
\left(
\bad
a & \epsilon_1 & 0 \\[0.2cm]
0 & b & 0 \\[0.2cm]
0 & \eta_3 & d 
\ea
\right) \Rightarrow 
m_\nu = -e^{-i \omega} \, \frac{v_u^2}{M} \, 
\left( 
\bad 
\frac{a^2}{X} & \frac{a \, \epsilon_1}{X} & 0 \\[0.2cm]
\cdot & \frac{\epsilon_1^2}{X} 
+ 2 \frac{b \, \eta_3}{Y} \, e^{i \omega} & 
\frac{b \, d}{Y} \, e^{i \omega}\\[0.2cm]
\cdot & \cdot & 0 
\ea 
\right)
~.
\ee
The $ee$ and the $\mu\tau$ elements are allowed by $L_\mu - L_\tau$ and,  
as it should, the additional non-zero entries are suppressed by the small 
breaking parameters. 
The expressions relevant for high and low energy $CP$ violation are  
\bea \label{eq:CP1} \D 
\varepsilon_X \simeq -\frac{1}{2 \pi} \, \frac{1}{Y/X - X/Y}
\, \frac{b}{a^2} \, \epsilon_1^2 \, \eta_3 \, \sin \omega 
~\mbox{ and } ~\\[0.3cm] \D 
{\rm Im} \left\{ h_{12} \, h_{23} \, h_{31} \right\} = 
\left(\frac{v_u^2}{M} \right)^6 \, 
2 \frac{a^4 \, b^3 \, d^2}{X^3 \, Y^3} \, 
\epsilon_1^2 \, \eta_3 \, \sin \omega
~.
\eea
The decay $\tau \to e \gamma$ is forbidden, whereas the branching ratio 
for $\mu \ra e \gamma$ ($\tau \ra \mu \gamma$) is proportional to 
$|a \, \epsilon_1 \, L_1|^2$ ($|d \, \eta_3 \, L_2 |^2$). 
In what regards these LFV decays, let us return to Eqs.~(\ref{eq:estimate}) 
and (\ref{eq:mdtmd}). 
Given the fact that $a$ is of order one, it is apparent that 
$|(\tilde{m}_D^\dagger \, L \, \tilde{m}_D)_{12}|^2 /v_u^4$ is of the 
order of $\epsilon_1^2 \, L_1^2 \sim 10 \, \epsilon_1^2$.  
From Eq.~(\ref{eq:estimate}) we see that for typical values of 
$\tilde m_S \simeq 200$ GeV and $\tan^2 \beta \simeq 10^2$, the 
branching ratio for $\mu \ra e \gamma$ is roughly 
given by $10^{-6} \, \epsilon_1^2$. 
This indicates small values of $\epsilon_1$, which however also 
decreases the decay asymmetry parameter $\varepsilon_1$, 
which is relevant for leptogenesis. With this crude estimate 
we can see that the requirement of successful leptogenesis makes 
the branching ratio of $\mu \ra e \gamma$ in general rather large, thereby 
snookering such scenarios. 
In principle one could let the heavy neutrino masses be extremely 
degenerate, so that the decay asymmetry is large even for small 
perturbative parameters, but this is regarded as fine-tuning. 
The underlying reason for the potentially too large branching ratios 
(for more model-independent analyzes, see for instance \cite{others}) 
is that the entries allowed by the 
symmetry in $m_D$ are all of order one. 
It is therefore a generic issue of the framework. 

\begin{center}
\begin{figure}[ht]
\epsfig{file=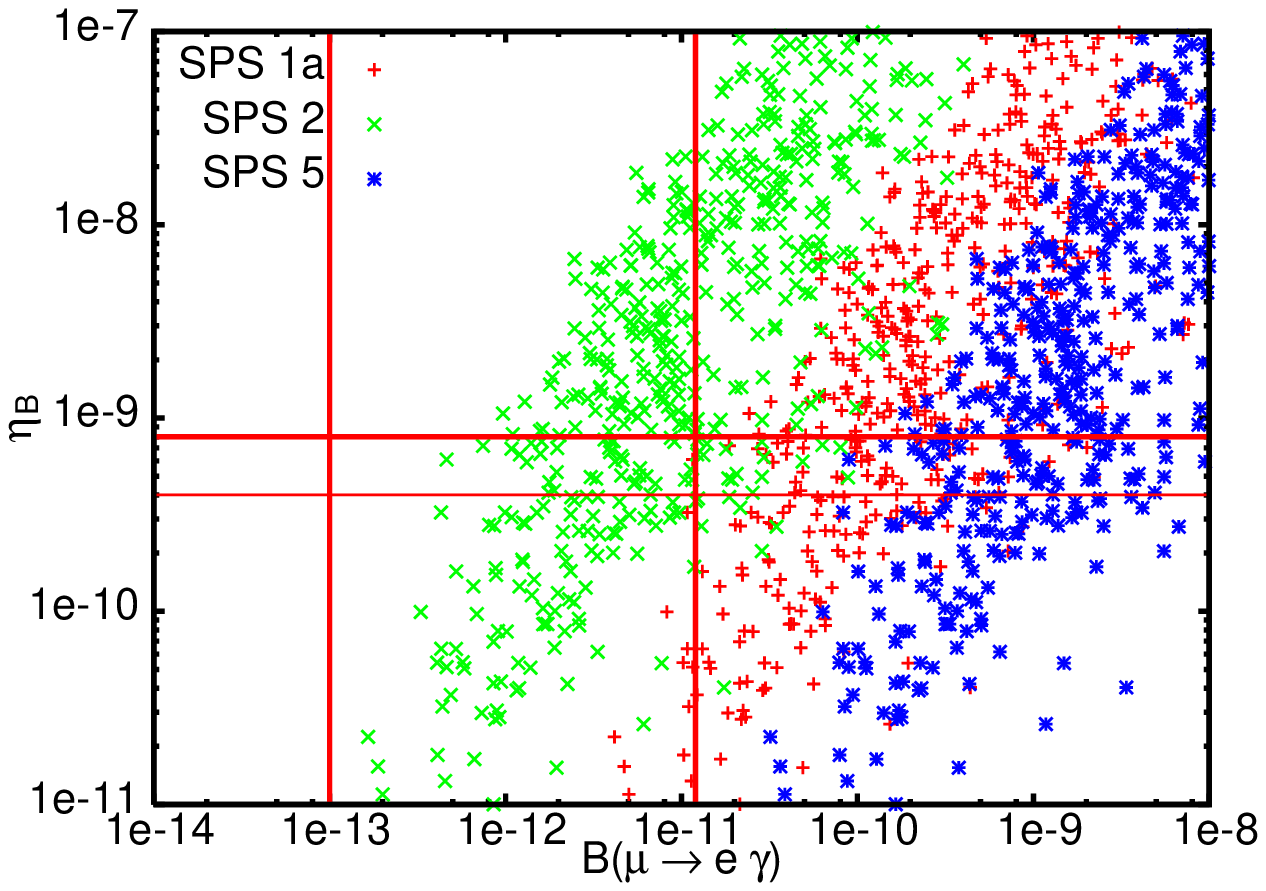,width=8.59cm,height=7.8cm}
\epsfig{file=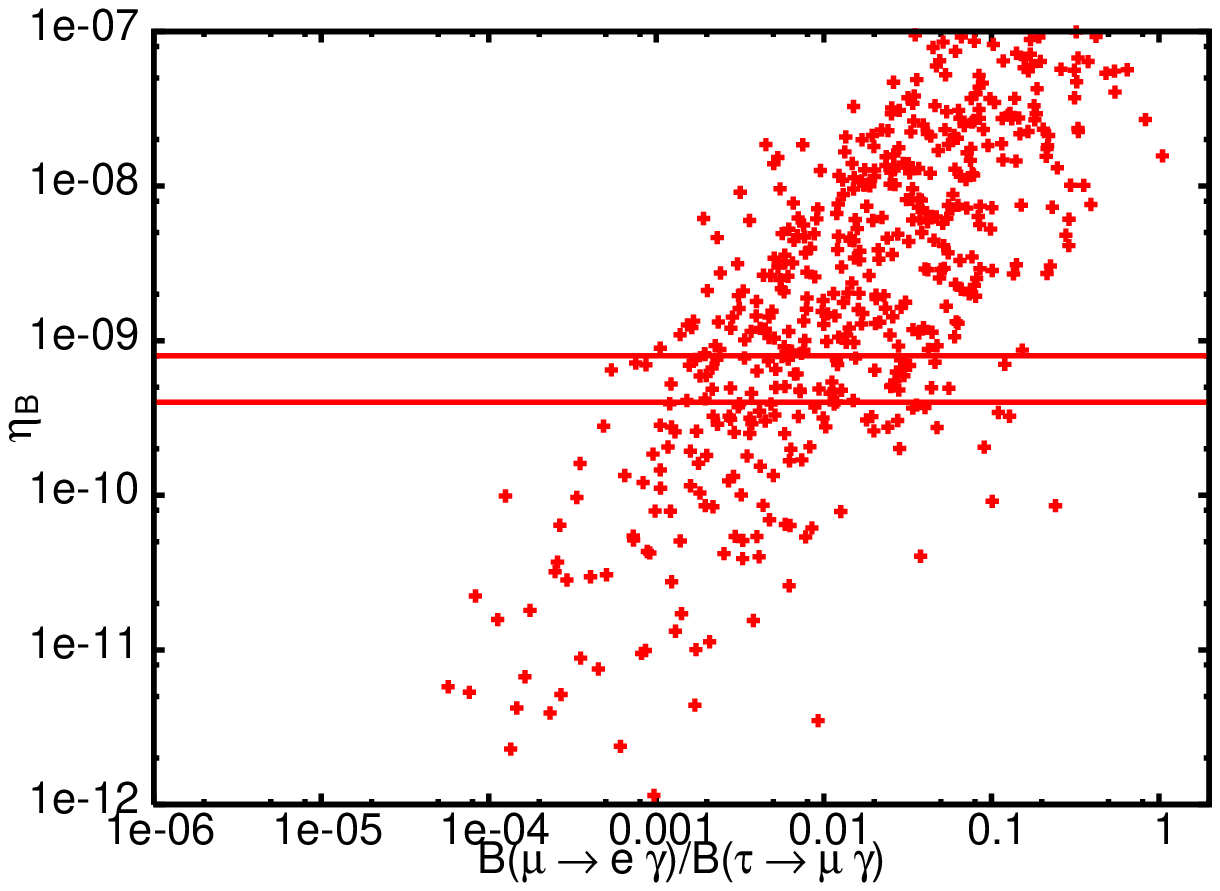,width=8.59cm,height=7.8cm}  \\
\epsfig{file=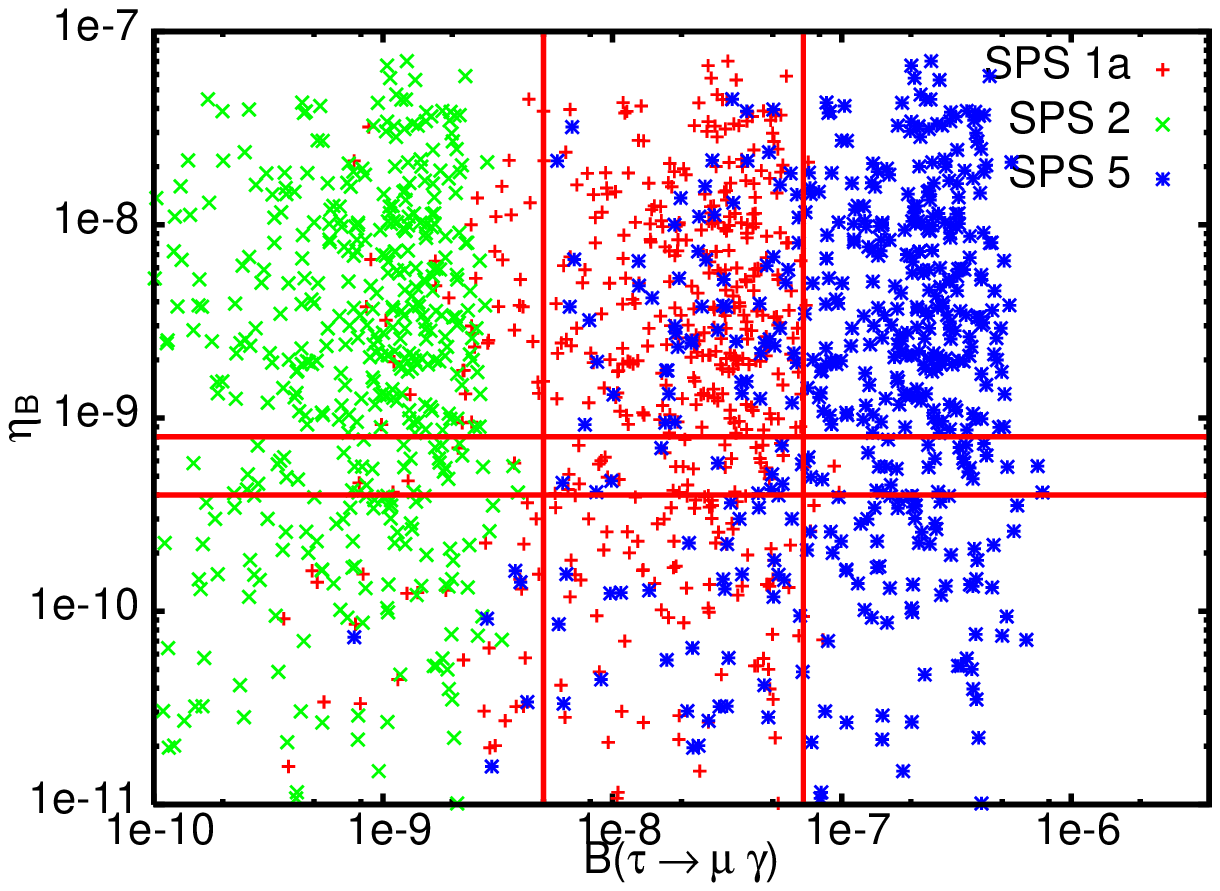,width=8.59cm,height=7.8cm}
\epsfig{file=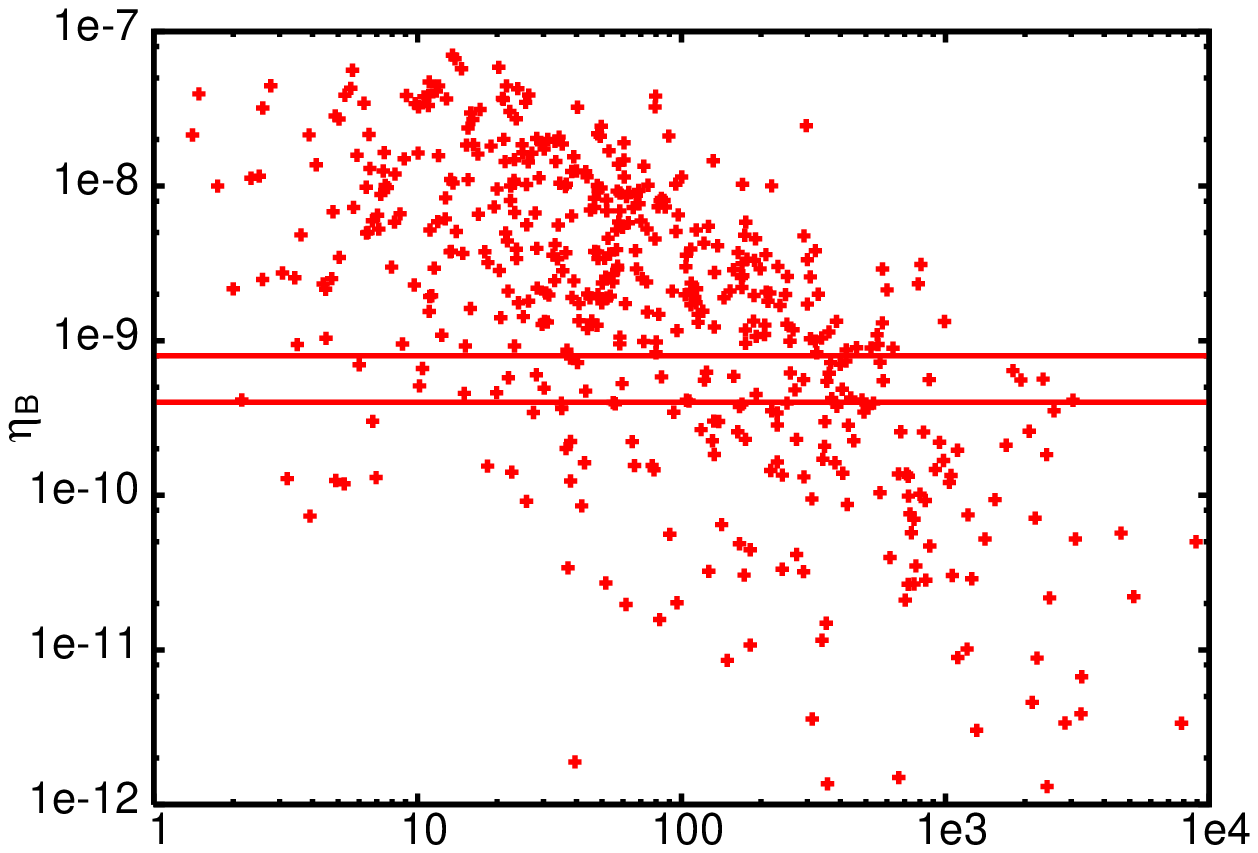,width=8.59cm,height=7.8cm}
\caption{\label{fig:ratios2}Magnitude of a LFV decay and the ratio of 
the non-zero branching ratios against the baryon asymmetry. The two upper 
plots are for Eq.~(\ref{eq:exam1}) and the two lower plots are for 
Eq.~(\ref{eq:exam2}).}
\end{figure}
\end{center}

We next perform a numerical search for successful parameters $a,b,d,X,Y$ 
(which are required to be of order one) 
and for the two perturbative parameters (which are required 
to be at least one order of magnitude smaller). 
The neutrino oscillation observables are required to lie 
within their 3$\sigma$ ranges from Ref.~\cite{thomas}. 
We also demand $1 - X/Y  \ge 0.1$ so that the heavy neutrinos 
are not too close in mass, i.e., Eq.~(\ref{e1H}) can still be used. 
We checked that the corrections to Eq.~(\ref{e1H}) are indeed subleading 
in this case. 
The upper left plot in Fig.~\ref{fig:ratios2} shows 
B$(\mu \ra e \gamma)$ against $\eta_B$ for the SPS benchmark 
points 1a, 2 and 5. 
It turns out that points 1a and 1b generate practically identical 
results, and also points 2 and 3 are indistinguishable. 
The results for point 4 lie between points 2 and 5. 
The correlation between $\eta_B$ and B$(\mu \ra e \gamma)$ is rather 
strong because both $\varepsilon_X$ and the branching ratio are proportional 
to $\epsilon_1^2$.
The upper right plot shows the ratio of the two non-zero 
branching ratios, which is below one for successful leptogenesis.  
We included the current and a 
future bound on the branching ratio and also indicated 
how many points lie in the range 
$4 \cdot 10^{-10} \le \eta_B \le 8 \cdot 10^{-10}$. Except for the SPS 
point 2, which includes TeV scale parameters, B$(\mu \ra e \gamma)$ 
is typically too large\footnote{We remark that 
point 5 leads to a too small Higgs mass anyway \cite{vempati}.}.  
As mentioned before, reducing the order of 
magnitude of the small perturbative parameters will strongly reduce $\eta_B$. 
A way to evade this problem is either to assume the SUSY parameters 
to be very large or to assume a breaking scheme of $L_\mu - L_\tau$ 
with zero B$(\mu \ra e \gamma)$. 

Such an example is\footnote{The remaining 
three cases with interesting correlations of the neutrino observables 
are found to be very fine-tuned, i.e., the numerical search for 
successful parameter values hardly finds any points.}  
\be\label{eq:exam2}
m_D = v_u \, 
\left(
\bad
a & 0 & \epsilon_2 \\[0.2cm]
0 & b & \epsilon_3 \\[0.2cm]
0 & 0 & d 
\ea
\right) \Rightarrow 
m_\nu = -e^{-i \omega} \, \frac{v_u^2}{M} \, 
\left( 
\bad 
\frac{a^2}{X} & 0 & \frac{a \, \epsilon_2}{X} \\[0.2cm]
\cdot & 0 & \frac{b \, d}{Y} \, e^{i \omega} \\[0.2cm]
\cdot & \cdot & \frac{\epsilon_2^2}{X} 
+ 2 \frac{d \, \epsilon_3}{Y} \, e^{i \omega}
\ea 
\right)
~.
\ee
This example has no decay $\mu \ra e \gamma$, and the 
branching ratio for 
$\tau \ra e \gamma$ ($\tau \ra \mu \gamma$) is proportional to 
$|a \, \epsilon_2 \, L_1|^2$ ($|b \, \epsilon_3 \, L_2 |^2$). 
$CP$ violation is governed by   
\bea\label{eq:CP2} \D
\varepsilon_X \simeq -\frac{1}{2 \pi} \, \frac{1}{Y/X - X/Y}
\, \frac{d}{a^2} \, \epsilon_2^2 \, \epsilon_3 \, \sin \omega 
~\mbox{ and } ~\\[0.3cm]
\D {\rm Im} \left\{ h_{12} \, h_{23} \, h_{31} \right\} = 
-\left(\frac{v_u^2}{M} \right)^6 \,
 2 \frac{a^4 \, d^3 \, b^2}{X^3 \, Y^3} \, 
\epsilon_2^2 \, \epsilon_3 \, \sin \omega
~.
\eea
The lower left plot of Fig.~\ref{fig:ratios2} shows 
B$(\tau \ra \mu \gamma)$ against $\eta_B$. 
We included the current ($6.8 \cdot 10^{-8}$ \cite{tmg}) and a 
future bound ($5 \cdot 10^{-9}$, see \cite{vempati}) 
on the branching ratio and also indicated 
how many points lie in the range 
$4 \cdot 10^{-10} \le \eta_B \le 8 \cdot 10^{-10}$. 
We see that $\tau \ra \mu \gamma$ lies in an observable range unless the 
SUSY masses are in the TeV range. 
The correlation between $\eta_B$ and the branching ratio is weaker 
than in the previous example, 
because $\varepsilon_X \propto \epsilon_3$ but 
B$(\tau \ra \mu \gamma) \propto \epsilon_3^2$.
The lower right plot shows the ratio of the two non-zero 
branching ratios, which is above one.

\begin{figure}[ht]
\begin{center}
\epsfig{file=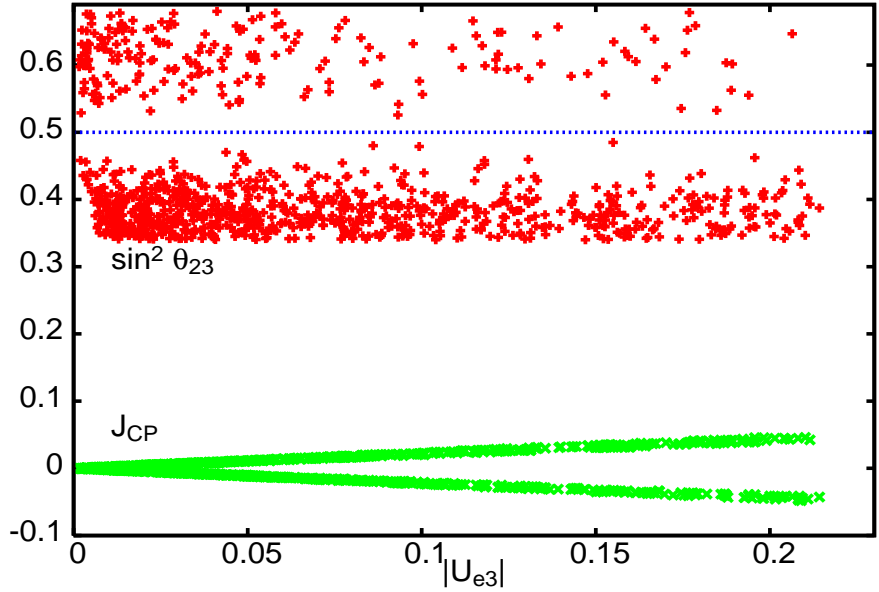,width=12cm,height=8cm}
\caption{\label{fig:obsA}Neutrino oscillation observables for 
Eq.~(\ref{eq:exam2}). Plotted is $|U_{e3}|$ against $J_{CP}$ and 
against $\sin^2 \theta_{23}$. Atmospheric neutrino mixing 
can not be exactly maximal 
and if $|U_{e3}|$ is large, $CP$ violation is also large. 
The results for Eq.~(\ref{eq:exam1}) are identical.}
\end{center}
\end{figure}

\begin{figure}[ht]
\begin{center}
\epsfig{file=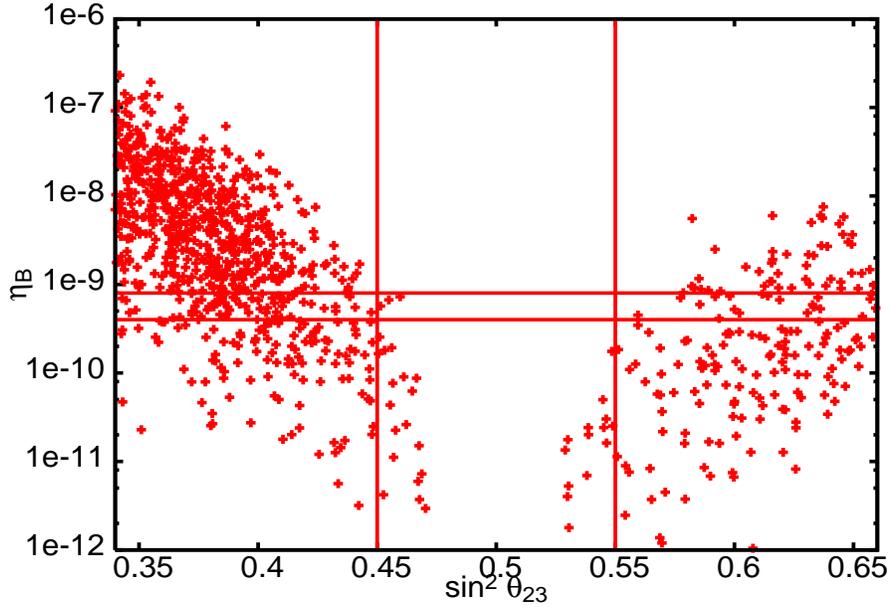,width=12cm,height=8cm}
\caption{\label{fig:obsA1}Atmospheric neutrino oscillation 
observable $\sin^2 \theta_{23}$ against the baryon asymmetry $\eta_B$ for 
Eq.~(\ref{eq:exam2}). 
The closer $\theta_{23}$ is to $\pi/4$, the smaller 
becomes $\eta_B$.}
\end{center}
\end{figure}

It is of course possible to diagonalize the mass matrices 
Eqs.~(\ref{eq:exam1}, \ref{eq:exam2}) and express the observables in 
terms of the parameters appearing in $m_D$ and $M_R$, 
but the resulting expressions are rather cumbersome 
and little insight is gained. We rather note that 
from the condition that the $e\tau$ and $\tau\tau$ entries 
(or the $e\mu$ and $\mu\mu$) vanish, one can obtain \cite{2zeros}
\be \label{eq:correlation}
\left| |U_{e3}| \, \cos \delta \, \tan 2 \theta_{23} \right| \simeq 
\frac{\dms}{2\dma} \, \sin 2 \theta_{12} 
~.
\ee 
Since $\dms/\dma \ll 1$, this expression means that 
$\theta_{23}$ can not be exactly 
maximal: $\sin^2 \theta_{23} \neq \frac 12$. 
Moreover, if $|U_{e3}|$ is sizable then $\cos \delta$ must be small, 
and therefore large $CP$ violation is expected in this case: 
$J_{CP} \simeq |U_{e3}|/4$. 
These features are nicely illustrated in Fig.~\ref{fig:obsA}, where we have 
plotted $|U_{e3}|$ against $J_{CP}$ and 
against $\sin^2 \theta_{23}$. Atmospheric neutrino mixing 
can not be exactly maximal 
and if $|U_{e3}|$ is large, $CP$ violation is also large. 
Identical results occur for Eq.~(\ref{eq:exam1}). 
Another interplay of variables occurs when $\theta_{23}$ is close to 
maximal. This implies again from Eq.~(\ref{eq:correlation}) 
that $\cos \delta$ is small and $J_{CP}$ is large. Large $J_{CP}$, 
in turn, implies from Eq.~(\ref{eq:CP2}) that the decay asymmetry is 
large, because both $\varepsilon_X$ and $J_{CP}$ are proportional to 
$\sin \omega$. Hence, the closer $\theta_{23}$ is to $\pi/4$, the smaller 
becomes $\eta_B$. This is illustrated in Fig.~\ref{fig:obsA1}. 
We indicated the values $\sin^2 \theta_{23} = 0.45$ and 0.55, which are 
the approximate lower and upper limits in order to still have 
successful leptogenesis. 

We stress here that both examples, Eqs.~(\ref{eq:exam1}, \ref{eq:exam2}), 
predict basically identical neutrino phenomenology, 
but differ dramatically in their predictions for the LFV decays.

\section{Summary and Conclusions}
A supersymmetric see-saw model obeying the flavor symmetry 
$L_\mu - L_\tau$ was analyzed. In the low energy sector this generates 
quasi-degenerate neutrinos, vanishing $\theta_{13}$ and maximal 
atmospheric neutrino mixing. With strict conservation of the symmetry 
both leptogenesis and LFV are not possible and in addition 
$\theta_{12}$ and the atmospheric $\Delta m^2$ is zero. 
Possibilities to break $L_\mu - L_\tau$ were considered and 
it was found that the most economic possibility is to include 
breaking only in $m_D$. Two small breaking parameters are 
required in order to allow for low energy $CP$ violation. 
Generation of the baryon asymmetry via leptogenesis is possible with 
heavy neutrino masses of similar size. 
We discussed how the breaking of the symmetry reflects in low energy 
observables, and in particular in the predictions for the 
LFV decays $\mu \ra e \gamma$, $\tau \ra e \gamma$ and $\tau \ra \mu \gamma$. 
Scenarios with indistinguishable neutrino phenomenology can lead to 
drastically different relations between the branching ratios.

\vspace{0.5cm}
\begin{center}
{\bf Acknowledgments}
\end{center}
This work was supported by the ``Deutsche Forschungsgemeinschaft'' in the 
``Sonderforschungsbereich 375 f\"ur Astroteilchenphysik'' 
(and T.O.~and W.R.) and under project number RO--2516/3--1 (W.R.).

\begin{table}\hspace{-1.5cm}
\begin{tabular}{|c|c|c|c|c|c|c|c|} \hline 
$m_D$ & $m_\nu$ & {\scriptsize $\mu \ra e \gamma$} & 
 {\scriptsize $\tau \ra e \gamma$} & 
  {\scriptsize $\tau \ra \mu \gamma$} & $J_{CP} $ & $\eta_B$ & Correlation 
\\ \hline \hline 
$\scriptsize \left( \bad 
a & \epsilon_1 & \epsilon_2  \\
0 & b & 0 \\ 
0 & 0 & d 
\ea \right)$ 
& 
$\scriptsize \left( \bad 
\times & \times & \times \\
\times & \times & \times \\ 
\times & \times & \times 
\ea \right)$ & $\times$ & $\times$ & $\times$ & $\times $ & $\times $ & 
------ \\ \hline 
$\scriptsize \left( \bad 
a & \epsilon_1 & 0  \\
0 & b & \epsilon_3 \\ 
0 & 0 & d 
\ea \right)$ & 
$\scriptsize \left( \bad 
\times & \times & 0 \\
\times & \times & \times \\ 
0 & \times & \times 
\ea \right)$ & $\times$ & $0$ & $\times$ & $\times $ & 0  & 
$ \ba \mbox{if QD:} \sin \alpha=0 \\ \Rightarrow \meff \simeq 
m_0 \ea $ \\ \hline 
$\scriptsize \left( \bad 
a & 0 & \epsilon_2  \\
0 & b & \epsilon_3 \\ 
0 & 0 & d 
\ea \right)$ & 
$\scriptsize \left( \bad 
\times & 0 & \times \\
0 & 0 & \times \\ 
\times & \times & \times 
\ea \right)$ & $0$ & $\times$ & $\times$  & $\times$ & $\times $ & 
$ \ba \mbox{QD; 
both orderings} \\
|U_{e3}| \simeq \frac{R}{2} \, |\frac{\cot 2\theta_{23} }{\cos \delta }|
\sin 2 \theta_{12} 
\ea $ \\ \hline 
$\scriptsize \left( \bad 
a & \epsilon_1 & 0  \\
\eta_1 & b & 0 \\ 
0 & 0 & d 
\ea \right)$ & 
$\scriptsize \left( \bad 
\times & \times & \times \\
\times & \times & \times \\ 
\times & \times & 0 
\ea \right)$ & $\times$ & $0$ & $0$  & 0 & 0 & 
$ \ba \mbox{ if QD: } \\ \scriptsize | 
s_{23}^2 \, \left( e^{2i\alpha} \, c_{12}^2 + s_{12}^2\right) 
+ e^{2i\left( \beta + \delta \right) }\,c_{23}^2|=0 \ea$ \\ \hline 
$\scriptsize \left( \bad 
a & \epsilon_1 & 0  \\
0 & b &  0\\ 
\eta_2 & 0 & d 
\ea \right)$ & 
$\scriptsize \left( \bad 
\times & \times & 0 \\
\times & \times & \times \\ 
0 & \times & 0 
\ea \right)$ & $\times$ & $\times$ & $0$  &  $\times$ & $\times $ & 
$\ba \mbox{QD; 
both orderings} \\
|U_{e3}| \simeq \frac{R}{2}  \, |\frac{\cot 2\theta_{23} }{\cos \delta }|
\sin 2 \theta_{12}
\ea $ \\ \hline 
$\scriptsize \left( \bad 
a &  \epsilon_1 & 0  \\
0 & b & 0 \\ 
0 & \eta_3 & d 
\ea \right)$ & 
$\scriptsize \left( \bad 
\times & \times & 0 \\
\times & \times & \times \\ 
0 & \times & 0
\ea \right)$ & $\times$ & $0$ & $\times$  &   $\times $ & $\times $ & 
$ \ba \mbox{QD; 
both orderings} \\
|U_{e3}| \simeq \frac{R}{2}  \, |\frac{\cot 2\theta_{23} }{\cos \delta }|
\sin 2 \theta_{12}
\ea $ \\ \hline 
$\scriptsize \left( \bad 
a & 0 & \epsilon_2  \\
\eta_1 & b & 0 \\ 
0 & 0 & d 
\ea \right)$ & 
$\scriptsize \left( \bad 
\times & 0 & \times \\
0 & 0 & \times \\ 
\times & \times & \times 
\ea \right)$ & $\times$ & $\times$ & $0$  &  $\times$ & $\times $ & 
$ \ba \mbox{QD; 
both orderings} \\
|U_{e3}| \simeq \frac{R}{2} \, |\frac{\cot 2\theta_{23} }{\cos \delta }|
\sin 2 \theta_{12} 
\ea $ \\ \hline 
$\scriptsize \left( \bad 
a & 0 & \epsilon_2  \\
0 & b & 0 \\ 
\eta_2 & 0 & d 
\ea \right)$ & 
$\scriptsize \left( \bad 
\times & \times & \times \\
\times & 0 & \times \\ 
\times & \times & \times 
\ea \right)$ & $0$ & $\times$ & $0$  & 0 & 0 & 
 $ \ba \mbox{ if QD: } \\ \scriptsize | 
c_{23}^2 \, \left( e^{2i\alpha} \, c_{12}^2 + s_{12}^2\right) 
+ e^{2i\left( \beta + \delta \right) }\,s_{23}^2|=0 \ea$ \\ \hline 
$\scriptsize \left( \bad 
a & 0 & \epsilon_2  \\
0 & b & 0 \\ 
0 & \eta_3  & d 
\ea \right)$ & 
$\scriptsize \left( \bad 
\times & 0 & \times \\
0 & \times & \times \\ 
\times & \times & \times 
\ea \right)$ & $0$ & $\times$ & $\times$  & $\times $ & 0 & 
$ \ba \mbox{if QD:} \sin \alpha=0 \\ \Rightarrow \meff \simeq 
m_0 \ea $ \\ \hline 
$\scriptsize \left( \bad 
a & 0 & 0  \\
\eta_1 & b & \epsilon_3 \\ 
0 & 0 & d 
\ea \right)$ & 
$\scriptsize \left( \bad 
\times & 0 & \times \\
0 & 0 & \times \\ 
\times & \times & \times 
\ea \right)$ & $\times$ & $\times$ & $\times$  & $\times $ & 0 & 
$ \ba \mbox{QD; 
both orderings} \\
|U_{e3}| \simeq \frac{R}{2} \, |\frac{\cot 2\theta_{23} }{\cos \delta }|
\sin 2 \theta_{12} 
\ea $ \\ \hline 
$\scriptsize \left( \bad 
a &  0 & 0   \\
0 & b & \epsilon_3 \\ 
\eta_2 & 0 & d 
\ea \right)$ & 
$\scriptsize \left( \bad 
\times & \times & \times \\
\times & 0 & \times \\ 
\times & \times & \times 
\ea \right)$ & $0$ & $\times$ & $\times$  & $\times $ & 0 & 
$ \ba \mbox{ if QD: } \\ \scriptsize | 
c_{23}^2 \, \left( e^{2i\alpha} \, c_{12}^2 + s_{12}^2\right) 
+ e^{2i\left( \beta + \delta \right) }\,s_{23}^2|=0 \ea$\\ \hline 
$\scriptsize \left( \bad 
a & 0 & 0  \\
0 & b & \epsilon_3 \\ 
0  & \eta_3 & d 
\ea \right)$ & 
$\scriptsize \left( \bad 
\times & 0 & 0 \\
0 & \times & \times \\ 
0 & \times & \times 
\ea \right)$ & $0$ & $0$ & $\times$  & 0 & 0 & 
ruled out by $m_\nu$\\ \hline 
$\scriptsize \left( \bad 
a & 0 & 0  \\
\eta_1 & b & 0 \\ 
\eta_2 & 0 & d 
\ea \right)$ & 
$\scriptsize \left( \bad 
\times & \times & \times \\
\times & 0 & \times \\ 
\times & \times & 0 
\ea \right)$ & $\times$ & $\times$ & $0$  &  $\times$ & $\times $ & 
$ \ba \mbox{ QD; only inverted } \\ \scriptsize | U_{e3}| \, \cos \delta 
\simeq \cot 2 \theta_{12} \, \cos 2 \theta_{23}  \ea$\\ \hline
 $\scriptsize \left( \bad 
a & 0 & 0  \\
\eta_1 & b & 0 \\ 
0 & \eta_3 & d 
\ea \right)$ & 
$\scriptsize \left( \bad 
\times & \times & \times \\
\times & \times & \times \\ 
\times & \times & 0 
\ea \right)$ & $\times$ & $0$ & $\times$  &  $\times $ & 0 & 
$ \ba \mbox{ if QD: } \\ \scriptsize | 
s_{23}^2 \, \left( e^{2i\alpha} \, c_{12}^2 + s_{12}^2\right) 
+ e^{2i\left( \beta + \delta \right) }\,c_{23}^2|=0 \ea$\\ \hline
$\scriptsize \left( \bad 
a & 0 & 0  \\
0 & b & 0 \\ 
\eta_2 & \eta_3 & d 
\ea \right)$ & 
$\scriptsize \left( \bad 
\times & \times & 0 \\
\times & \times & \times \\ 
0 & \times & 0 
\ea \right)$ & $\times$ & $\times$ & $\times$  & $\times $ & 0 & 
$\ba \mbox{QD; 
both orderings} \\
|U_{e3}| \simeq \frac{R}{2}  \, |\frac{\cot 2\theta_{23} }{\cos \delta }|
\sin 2 \theta_{12}
\ea $ \\ \hline
\end{tabular}
\caption{\label{tab:break}Dirac mass matrices with two non-zero 
breaking parameters, the resulting low energy mass matrix $m_\nu$, 
the implications for $\ell_j \ra \ell_i \gamma$, for 
low energy $CP$ violation, 
for $\eta_B$, and a correlation of the neutrino observables resulting 
from the form of $m_\nu$. QD means quasi-degenerate neutrinos with a common 
mass scale $m_0$ and $R$ is defined as $\dms/\dma$.}
\end{table}

\end{document}